\documentclass[letterpaper, 10 pt, conference]{ieeeconf}  

\IEEEoverridecommandlockouts                             
\usepackage[noadjust]{cite}
\usepackage[utf8]{inputenc} 
\usepackage[T1]{fontenc} 
\usepackage{hyperref}       
\usepackage{url}            
\usepackage{booktabs}      
\usepackage{amsfonts}      
\usepackage{nicefrac}      
\usepackage{microtype}    
\usepackage{algorithmic}
\usepackage[ruled,vlined]{algorithm2e}
\usepackage{amsmath}
\usepackage{graphicx}
\usepackage{multicol}
\usepackage[font=footnotesize, labelfont=bf]{caption}

\usepackage{bbm}

\usepackage{dblfloatfix}
\usepackage{subcaption,graphicx}

\usepackage{hyperref}
\hypersetup{colorlinks=true,linkcolor=blue,urlcolor=blue}

\usepackage[]{algorithm2e}
\SetKwInput{KwInput}{Input \hspace*{0.3em}}
\SetKwInput{KwOutput}{Output}

\usepackage{tikz}

\usepackage{amssymb}
\usepackage{pgfplots}
\usepackage{pgfmath}
\usepgfplotslibrary{patchplots}
\usetikzlibrary{patterns, positioning, arrows}

\pgfmathsetmacro\sprayRadius{.75pt}
\pgfmathsetmacro\sprayPeriod{.8cm}

\usepackage{chngcntr}
\counterwithin{paragraph}{section}

\usepackage{hyperref}

\title{\LARGE \bf
Score-Based Data Generation for EEG Spatial Covariance Matrices: Towards Boosting BCI Performance
}

\author{Ce~Ju$^{1}$, Reinmar Josef Kobler$^{2}$ and Cuntai~Guan$^{1}$~\IEEEmembership{Fellow, IEEE}
\thanks{
$^{1}$\, Ce Ju and Cuntai Guan are with the S-Lab and School of Computer Science and Engineering, Nanyang Technological University, 50 Nanyang Avenue, Singapore (emails: \{juce0001,ctguan\}@ntu.edu.sg).}
\thanks{
$^{2}$\, Reinmar Josef Kobler is with RIKEN Artificial Intelligence Project, Tokyo, and Advanced Telecommunications Research Institute International, Kyoto, Japan (email: reinmar.kobler@atr.jp).
        }
}

\begin{document}

\maketitle
\thispagestyle{empty}
\pagestyle{empty}

\begin{abstract}

The efficacy of Electroencephalogram (EEG) classifiers can be augmented by increasing the quantity of available data. 
In the case of geometric deep learning classifiers, the input consists of spatial covariance matrices derived from EEGs. 
In order to synthesize these spatial covariance matrices and facilitate future improvements of geometric deep learning classifiers, we propose a generative modeling technique based on state-of-the-art score-based models. 
The quality of generated samples is evaluated through visual and quantitative assessments using a left/right-hand-movement motor imagery dataset.
The exceptional pixel-level resolution of these generative samples highlights the formidable capacity of score-based generative modeling. 
Additionally, the center (Fréchet mean) of the generated samples aligns with neurophysiological evidence that event-related desynchronization and synchronization occur on electrodes C3 and C4 within the Mu and Beta frequency bands during motor imagery processing.
The quantitative evaluation revealed that 84.3\% of the generated samples could be accurately predicted by a pre-trained classifier and an improvement of up to 8.7\% in the average accuracy over ten runs for a specific test subject in a holdout experiment.
\end{abstract}

\section{INTRODUCTION}

Recently, deep learning techniques have been extensively adopted for classifying electroencephalography (EEG) data~\cite{schirrmeister2017deep}.
Despite this, a significant obstacle persists in the limited availability of training data~\cite{ju2020federated}.
To circumvent this constraint, researchers have turned to generative modeling, a rapidly evolving field in machine learning, to generate synthetic EEG time series through a process known as data augmentation~\cite{hartmann2018eeg}.  
This technique involves the creation of plausible samples that were not present in the original dataset, thereby expanding the training data with "unseen" examples.

To tackle the non-Euclidean characteristics inherent in EEG signals, researchers have been exploring the use of geometric deep learning methods to classify EEG signals in brain-computer interfaces (BCIs)~\cite{ju2022tensor,ju2022deep,ju2022graph,kobler_neurips22,pan2022matt,wilson2022deep}.
These techniques involve the application of second-order neural networks on matrices, known as spatial covariance matrices (SCMs), which are derived from EEG signals. 
It is noteworthy that these SCMs possess a wealth of discriminative information, including the variance of signals recorded by individual channels and the coherence between signals recorded by neighboring channels~\cite{ju2022graph}.
As a result, the development of generative models for SCMs with neurocognitive relevance presents a promising approach to enhancing the efficacy of geometric deep learning classifiers, ultimately delivering tangible advantages to BCI research.

In this study, we generate SCMs utilizing a cutting-edge generative modeling technique known as score-based generative modeling~\cite{song2019generative,song2020improved}.
Score-based generative models generate samples from noise by introducing a gradual increase in noise to the data, which is then undone through the estimation of the score function, which represents the gradient of the log-density function relative to the data.
This noise perturbation can be described as a forward diffusion process modeled by a stochastic differential equation (SDE)~\cite{song2021scorebased}.
This approach has been demonstrated to be successful in generating images, audio, and point clouds. 
In contrast to three-channel RGB images, which have pixel light intensities that range from 0 to 255, SCMs are generally preprocessed as multiple-channel square matrices that possess both symmetric and positive semidefinite properties and are decimal entities. 
We evaluate our approach using the Korea University (KU) dataset, which is the largest EEG-BCI dataset for two-class motor imagery classification.

\begin{figure}[!h]
\center
\includegraphics[width=\linewidth]{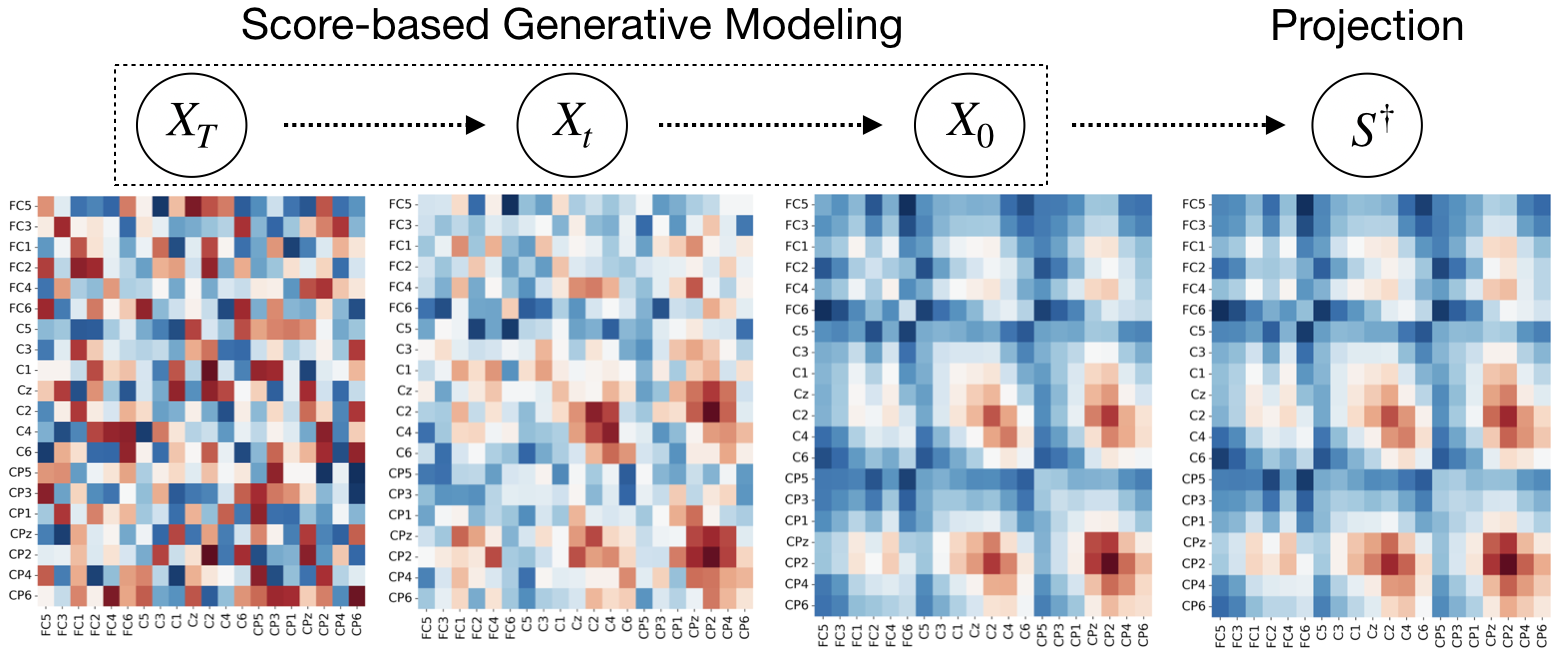}
\caption{The procedure of sampling can be depicted in the subsequent manner: Initially, we possess a noise matrix, represented as $X_T$. By employing the score-based generative modeling technique, we synthesize the spatial covariance matrix, $X_0$, with an intermediate state denoted as $X_t$.
Owing to the acquired knowledge, the generated SCM $X_0$ is approximated as being nearly symmetric positive definite (SPD).
In order to counterbalance numerical inconsistencies, we ensure the projection of $X_0$ as SPD by imposing a threshold upon all eigenvalues, denoted as $\epsilon = 1e-4$.
The arrangement of SCM channels proceeds sequentially from beginning to end as follows: FC-5/3/1/2/4/6, C-5/3/1/z/2/4/5, and CP-5/3/1/z/2/4/6.
~\label{fig:diagram}
}
\end{figure}

\section{METHODOLOGY}
In this section, we propose a two-step process for generating SCMs utilizing score-based generative modeling. 
The mathematical foundations of score-based generative modeling and projection are presented in Section~\ref{A1},~\ref{A2}, and~\ref{sec:projection}.
The process of sampling is depicted in Figure~\ref{fig:diagram}. 

\begin{itemize}
\item \textbf{Score-based Generative Modeling}: 
During the training process, unprocessed EEG signals undergo filtration and segmentation in both frequency and temporal domains, employing methodologies delineated in~\cite{ju2022tensor,ju2022graph}.
Explicitly, a collection of bandpass filters (i.e., Chebyshev Type II filters) is utilized to dissect the EEGs into multiple-frequency passbands. Subsequently, a temporal segmentation scheme is executed to partition the EEGs into diminutive segments, with or without overlap.
For a segment within $T$ duration $X \in \mathbb{R}^{n_C \times n_T}$, the spatial covariance matrix is denoted as $S:=X \cdot X^{\top} \in \mathbb{R}^{n_C \times n_C}$, where $n_C$ and $n_T$ represent the quantity of channels and timestamps, correspondingly.
In the terminal procedure of this phase, SCMs undergo scaling by division with their respective $\ell_2-$norms, indicated as $\bar{S}:= S/||S||_{\ell_2}$.
Utilizing score-based generative modeling, the unknown prior distribution $p_{data}(S)$ is approximated through score matching, generating samples within specific frequency bands and temporal intervals of EEGs, employing either Langevin dynamics or time-reversal SDEs.
The model is concurrently fitted for all frequency bands.
During the sampling process, the generated samples consist of $n_C \times n_C$ matrices, albeit generally lacking symmetry and positivity.

\item \textbf{Projection}:
In this step, we approximate the generated matrices to preserve symmetry and positivity.
Specifically, suppose a generated matrix $X \in \mathbb{R}^{n_C \times n_C}$, then the projected SCM as follows, 
\[
S^{\dagger} := \sum_{i=1}^{n_C} \max\{\lambda_i, \epsilon\} u_i  u_i^{\top},
\] 
where $\epsilon>0$ is a preset threshold, eigenvalues $\{\lambda_i\}_{i=1}^{n_C}$ and corresponding orthonormal eigenvectors $\{u_i\}_{i=1}^{n_C}$ are crafted from symmetric matrix $\frac{1}{2}(X + X^{\top})$.
\end{itemize}

\begin{figure*}[!h]
\center
\begin{subfigure}{0.5\textwidth}
\includegraphics[width=\linewidth]{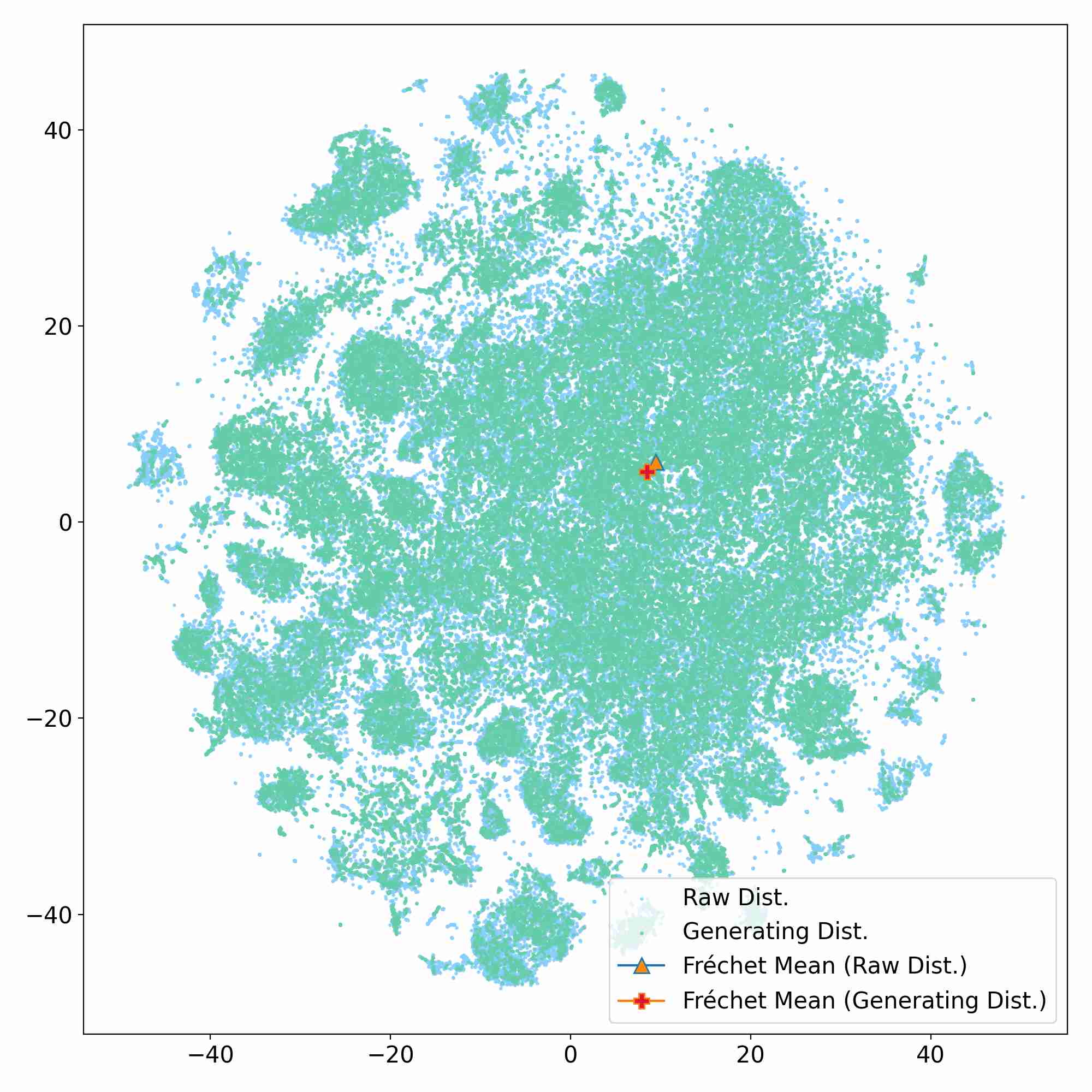}
\caption{Raw/Generating Dist. for all nine frequency bands}~\label{fig:a}
\end{subfigure}\hspace*{\fill}
\begin{subfigure}{0.5\textwidth}
\includegraphics[width=\linewidth]{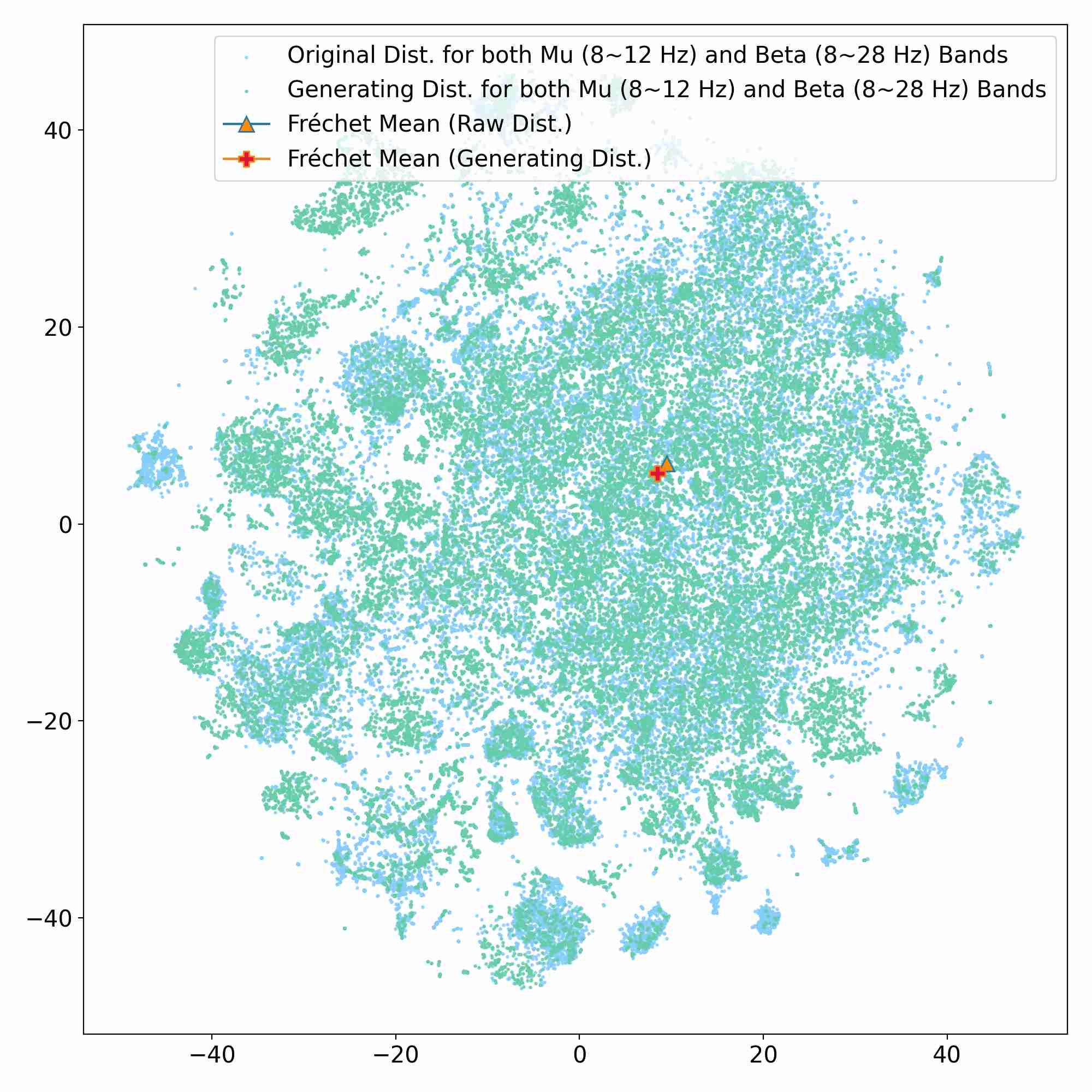}
\caption{Raw/Generating Dist. for the Mu and Beta frequency bands.}~\label{fig:b}
\end{subfigure}

\medskip

\begin{subfigure}{0.5\textwidth}
\includegraphics[width=\linewidth]{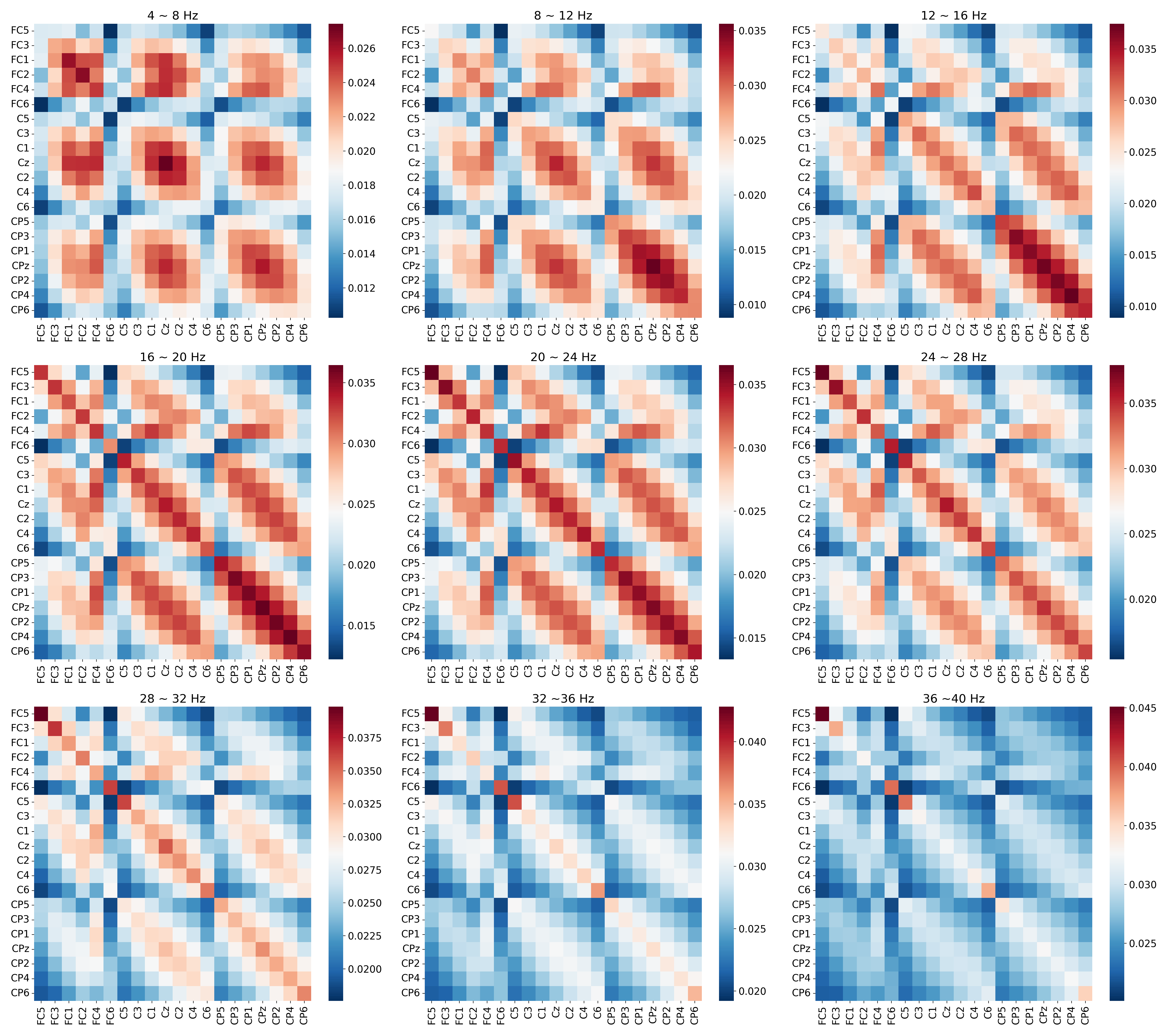}
\caption{Fréchet Mean of the raw dataset. (Triangle sign in Subfigure~\ref{fig:a})}~\label{fig:c}
\end{subfigure}\hspace*{\fill}
\begin{subfigure}{0.5\textwidth}
\includegraphics[width=\linewidth]{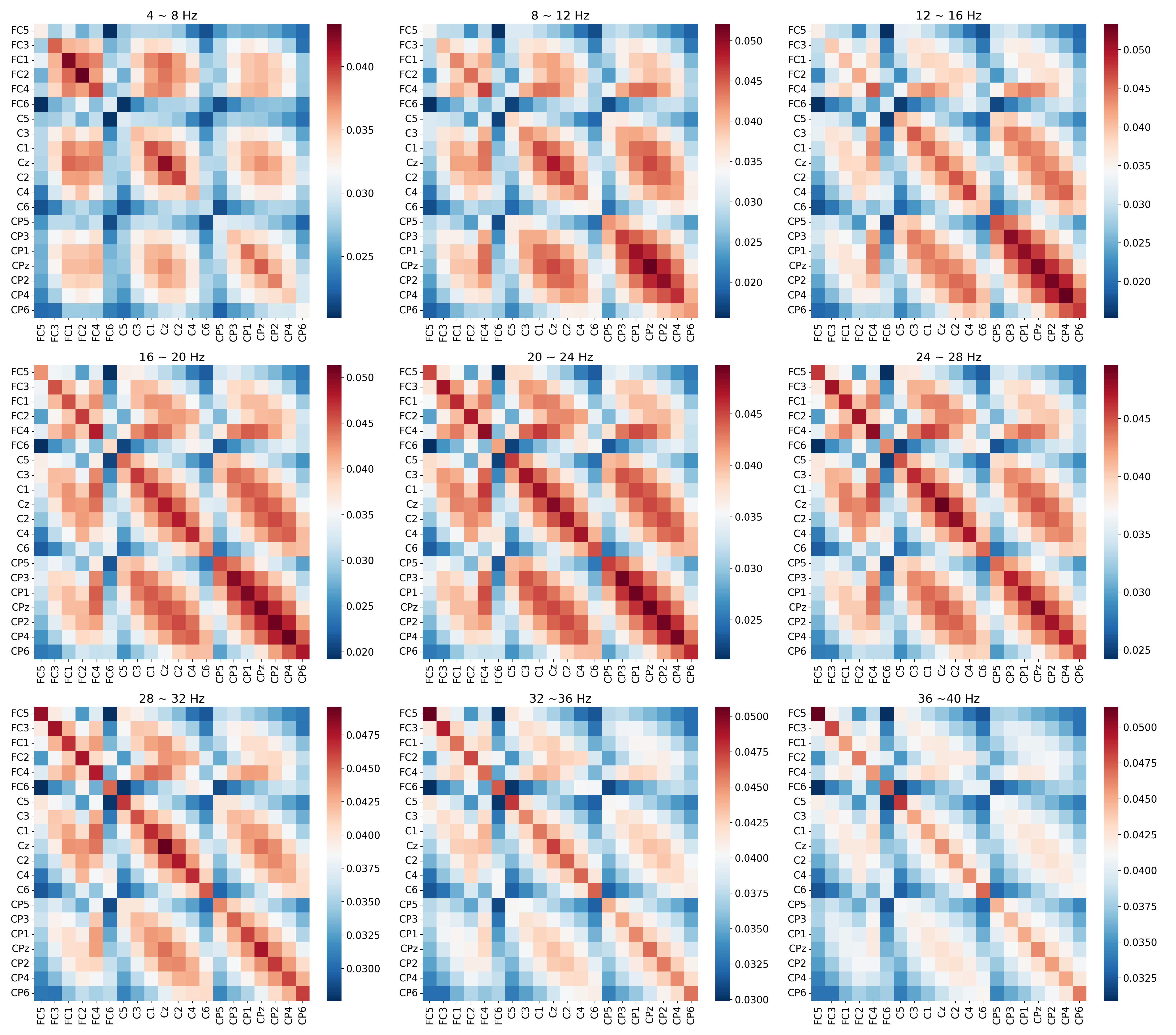}
\caption{Fréchet Mean of the generating dataset. (Cross sign in Subfigure~\ref{fig:a})}~\label{fig:d}
\end{subfigure}
\caption{\textbf{Subfigures~\ref{fig:a} and~\ref{fig:b}}: Illustration of raw and generating distributions of the 2-dimensional projection of covariance matrices:
Each 2-dimensional point in the figure is projected from its associated $20\times 20$ covariance matrices (i.e., a 400-dimensional tensor) using t-SNE.
There are 151,200 points (i.e., 9 frequency bands $\times$ 8400 trials $\times$ raw/generating options) in Subfigure~\ref{fig:a} and 84,000 points (i.e., 5 frequency bands $\times$ 8400 trials $\times$ raw/generating options) in Subfigure~\ref{fig:b}. 
\\
\textbf{Subfigures~\ref{fig:c} and~\ref{fig:d}}: Illustration of Fréchet means of covariance matrices for the nine frequency bands, 4 $\sim$ 8 Hz, 
8 $\sim$ 12 Hz, 
12 $\sim$ 16 Hz, 
16 $\sim$ 20 Hz, 
20 $\sim$ 24 Hz, 
24 $\sim$ 28 Hz, 
28 $\sim$ 32 Hz,
32 $\sim$ 36 Hz,
and 36 $\sim$ 40 Hz.
~\label{fig:1}
}
\end{figure*}


\begin{figure*}[!h]
\center
\begin{subfigure}{0.5\textwidth}
\includegraphics[width=\linewidth]{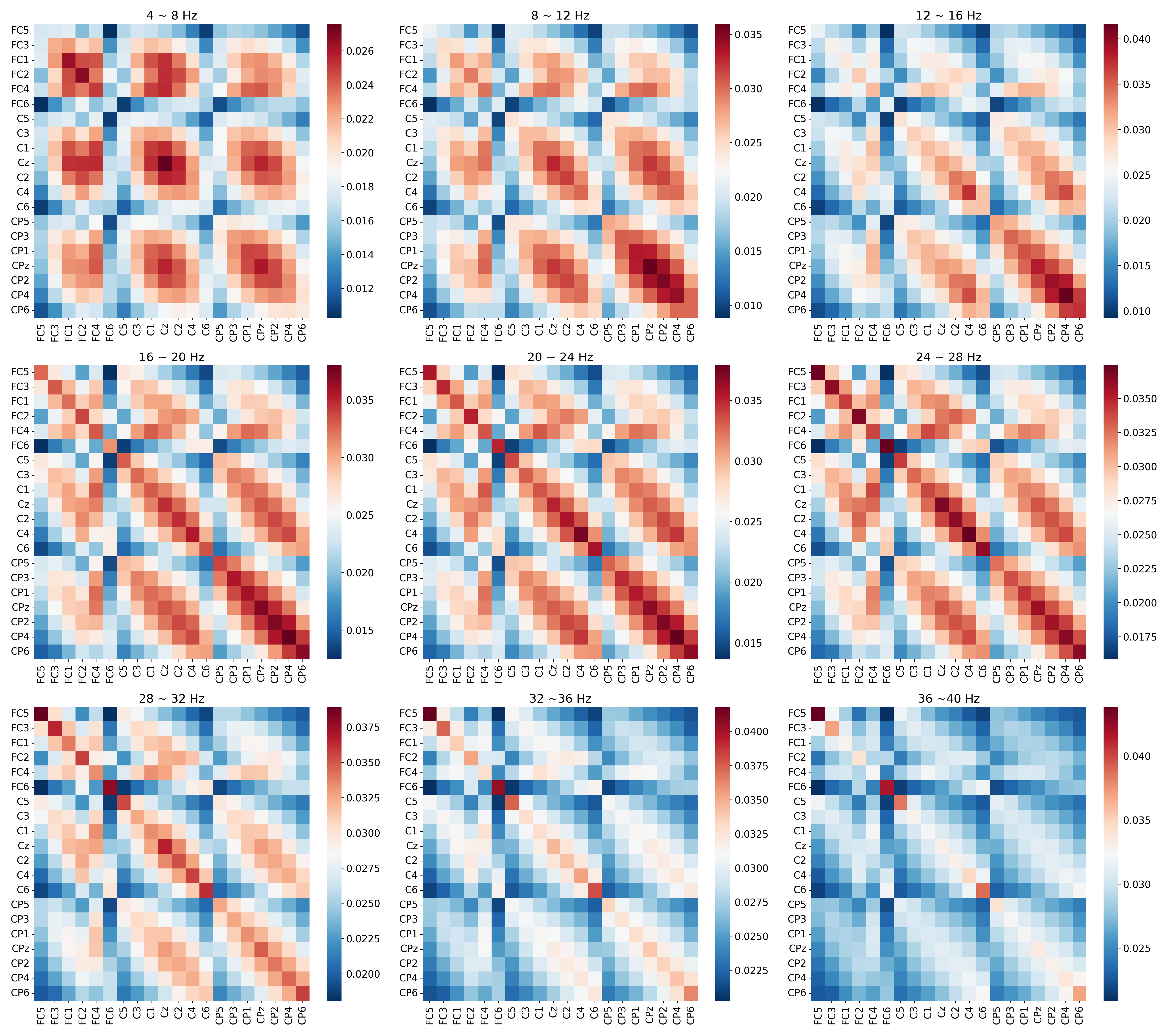}
\caption{Fréchet Mean of the \textbf{left}-hand-class trials in the raw dataset.}~\label{fig:e}
\end{subfigure}\hspace*{\fill}
\begin{subfigure}{0.5\textwidth}
\includegraphics[width=\linewidth]{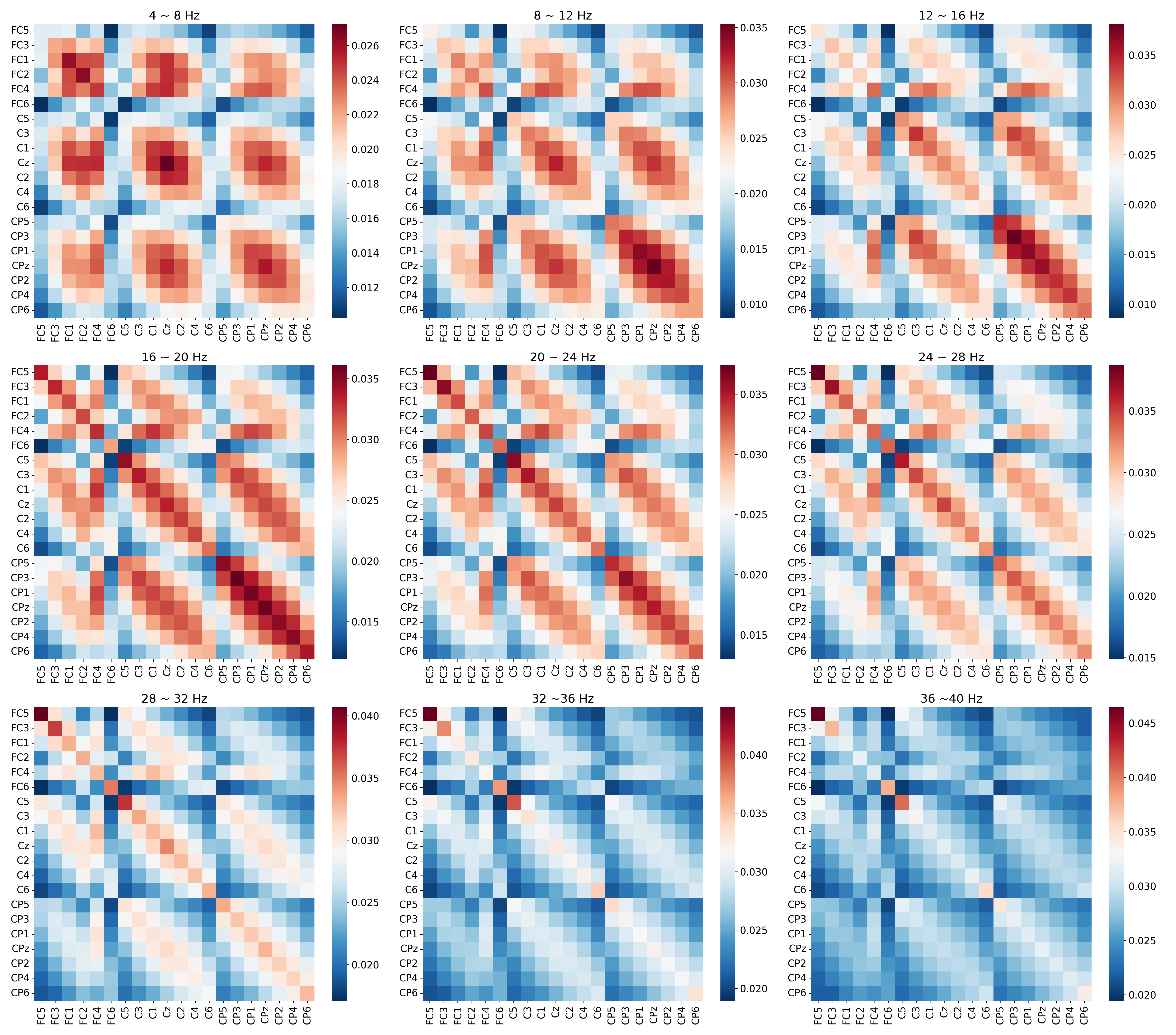}
\caption{Fréchet Mean of the \textbf{right}-hand-class trials in the raw dataset.}~\label{fig:f}
\end{subfigure}

\medskip

\begin{subfigure}{0.5\textwidth}
\includegraphics[width=\linewidth]{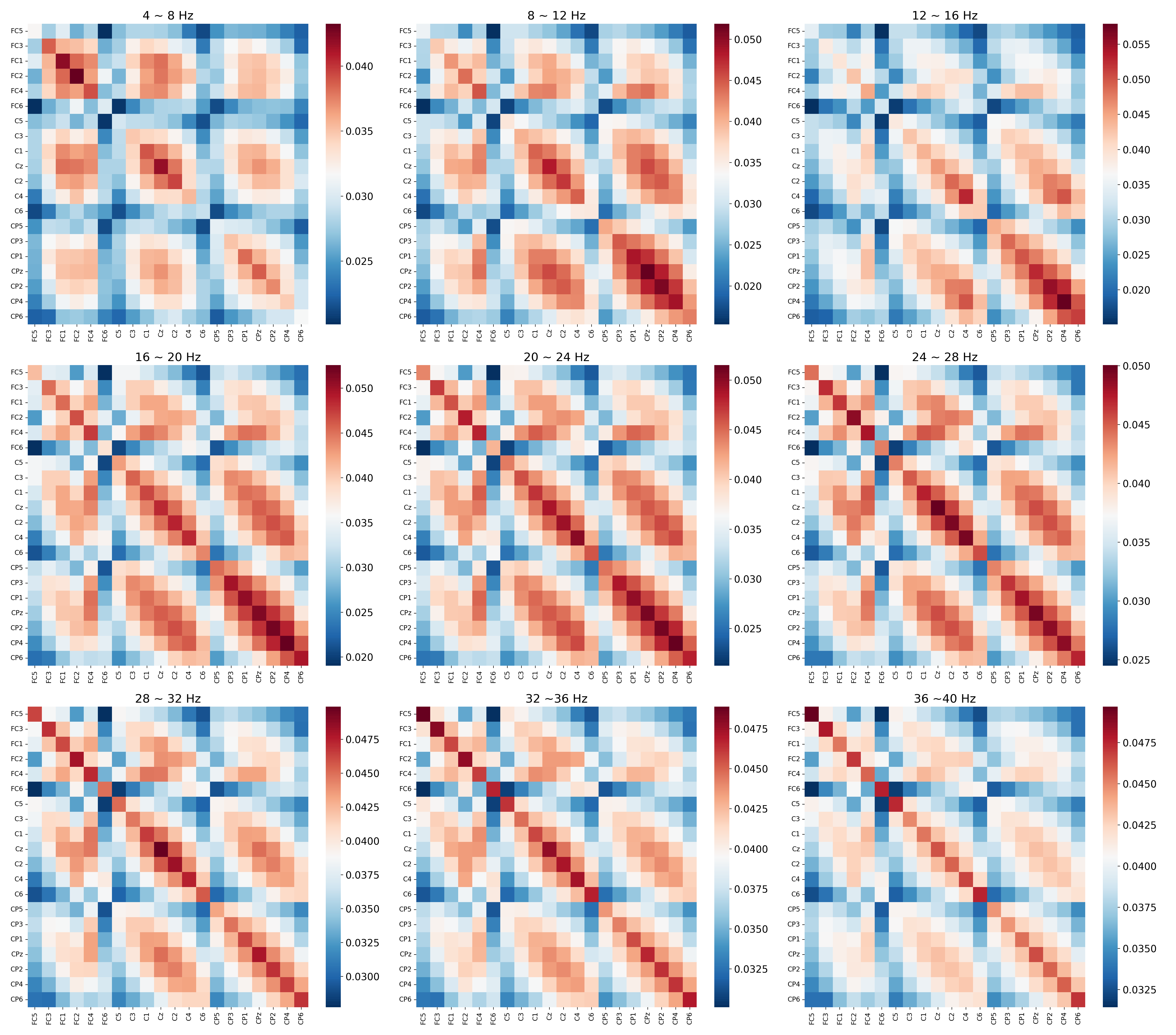}
\caption{Fréchet Mean of the \textbf{left}-hand-class trials in the generating dataset.}~\label{fig:g}
\end{subfigure}\hspace*{\fill}
\begin{subfigure}{0.5\textwidth}
\includegraphics[width=\linewidth]{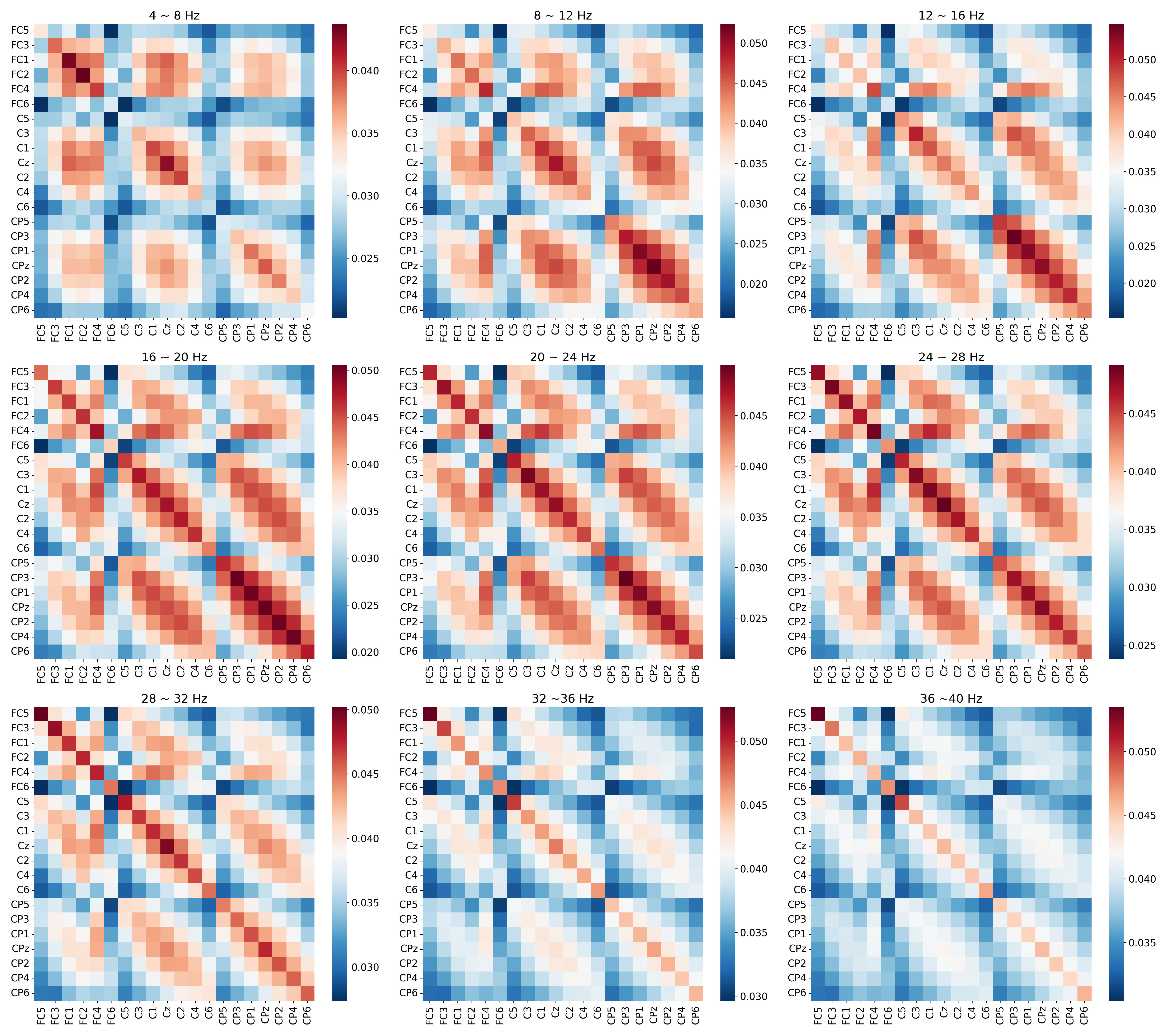}
\caption{Fréchet Mean of the \textbf{right}-hand-class trials in the generating dataset.}~\label{fig:h}
\end{subfigure}
\caption{Illustration of Fréchet means of covariance matrices within the nine frequency bands for the \textbf{left} and \textbf{right}-hand classes.
The highlight entities of SCMs in subfigures~\ref{fig:e} and~\ref{fig:g} (Mu and Beta bands) locate in the regions of FC4, C4, and CP4 over the scalp, while those in subfigures~\ref{fig:f} and~\ref{fig:h} fall in the regions of FC3, C3, and CP3.
~\label{fig:2}
}
\end{figure*}

\begin{figure*}[!ht]
\center
\begin{subfigure}{0.33\textwidth}
\includegraphics[width=\linewidth]{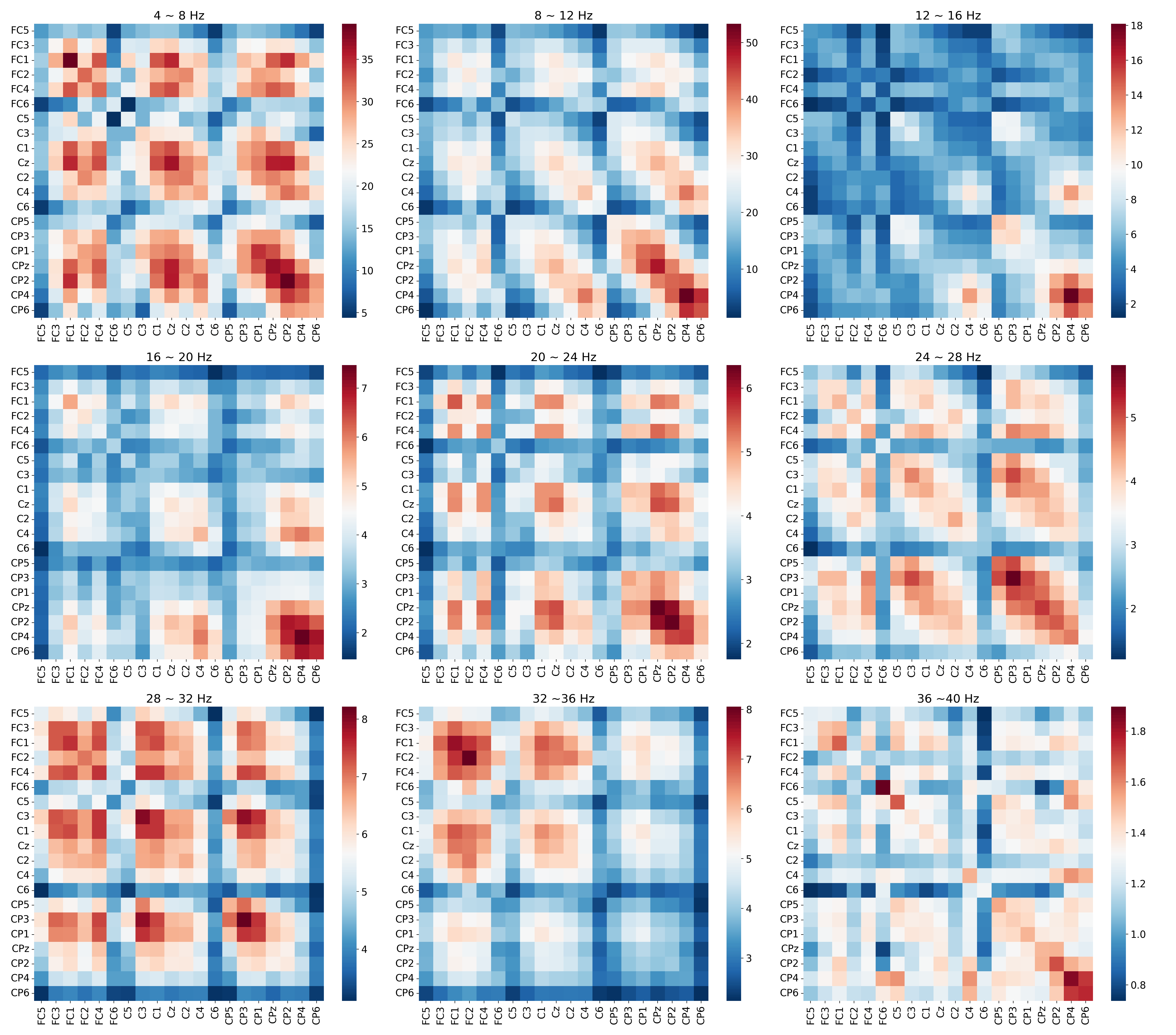}
\caption{Real SCMs, \textbf{left}-hand (Trial No.27 of Subject No.1)}~\label{fig:i}
\end{subfigure}\hspace*{\fill}
\begin{subfigure}{0.33\textwidth}
\includegraphics[width=\linewidth]{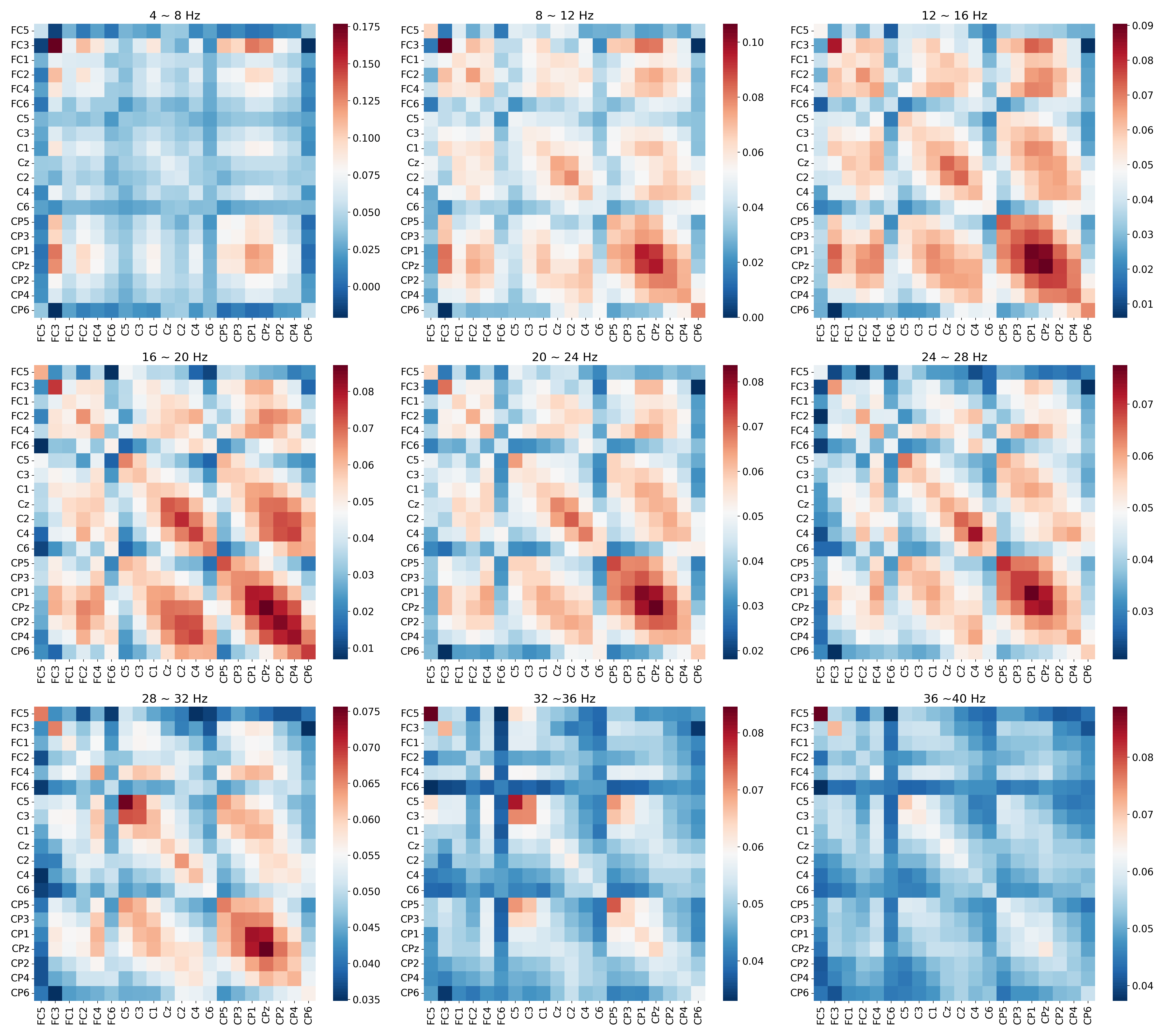} 
\caption{Generated SCMs, \textbf{left}-hand (Sample 1)}~\label{fig:j}
\end{subfigure}\hspace*{\fill}
\begin{subfigure}{0.33\textwidth}
\includegraphics[width=\linewidth]{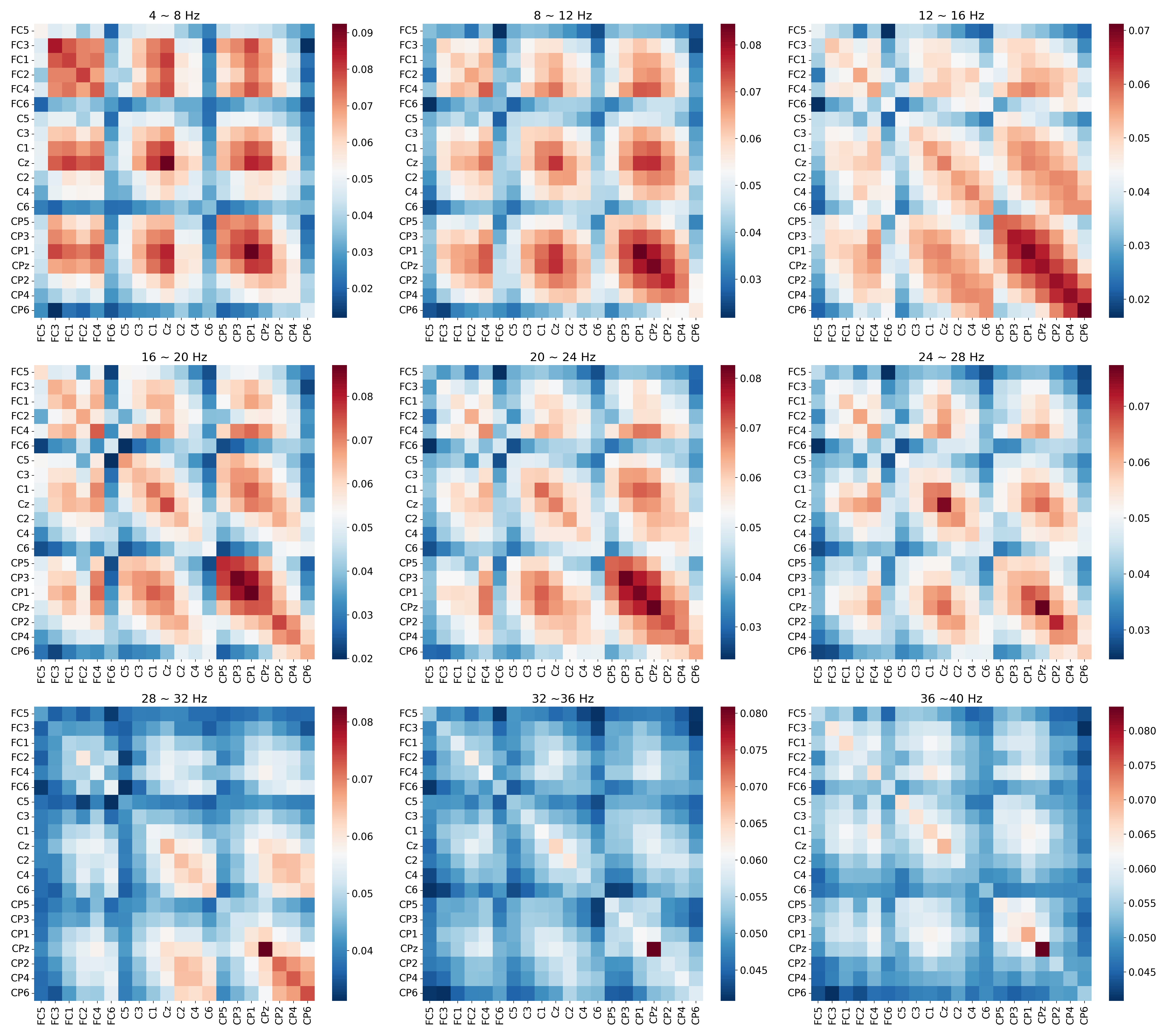}
\caption{Generated SCMs, \textbf{left}-hand (Sample 2)}~\label{fig:k}
\end{subfigure}
\medskip
\begin{subfigure}{0.32\textwidth}
\includegraphics[width=\linewidth]{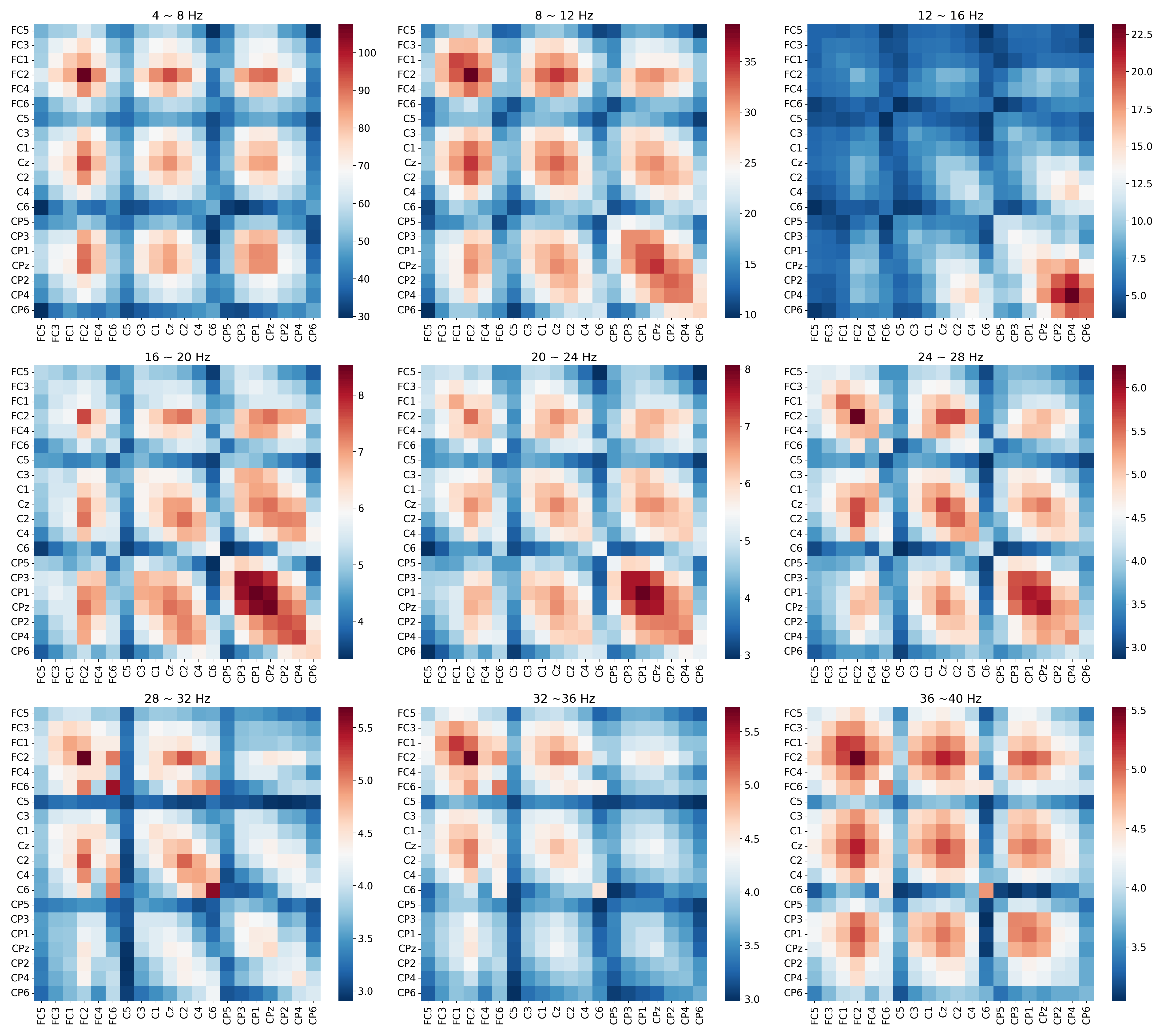}
\caption{Real SCMs, \textbf{right}-hand (Trial No.8 of Subject No.1)}~\label{fig:l}
\end{subfigure}\hspace*{\fill}
\begin{subfigure}{0.32\textwidth}
\includegraphics[width=\linewidth]{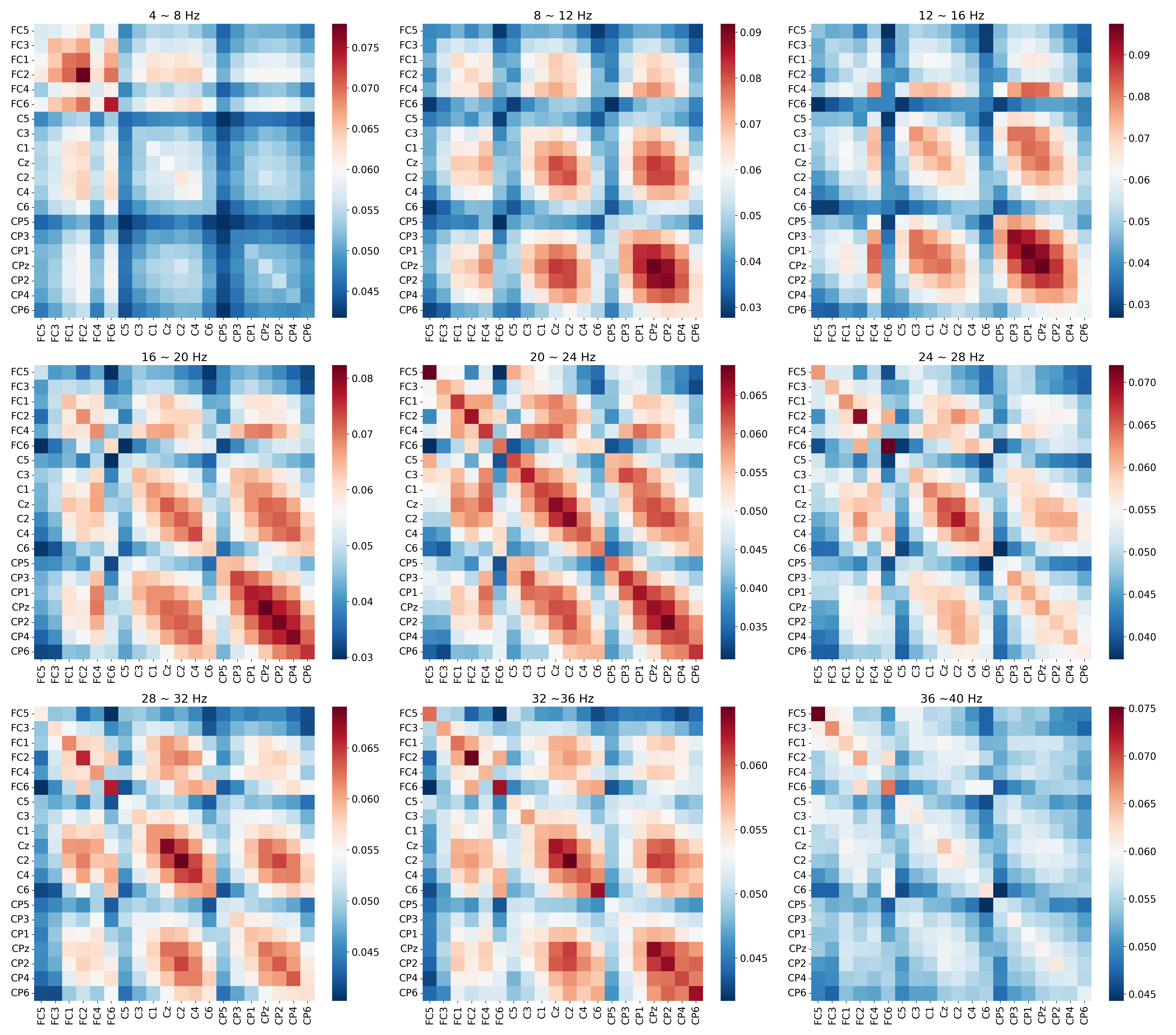}
\caption{Generated SCMs, \textbf{right}-hand (Sample 1)}~\label{fig:m}
\end{subfigure}\hspace*{\fill}
\begin{subfigure}{0.32\textwidth}
\includegraphics[width=\linewidth]{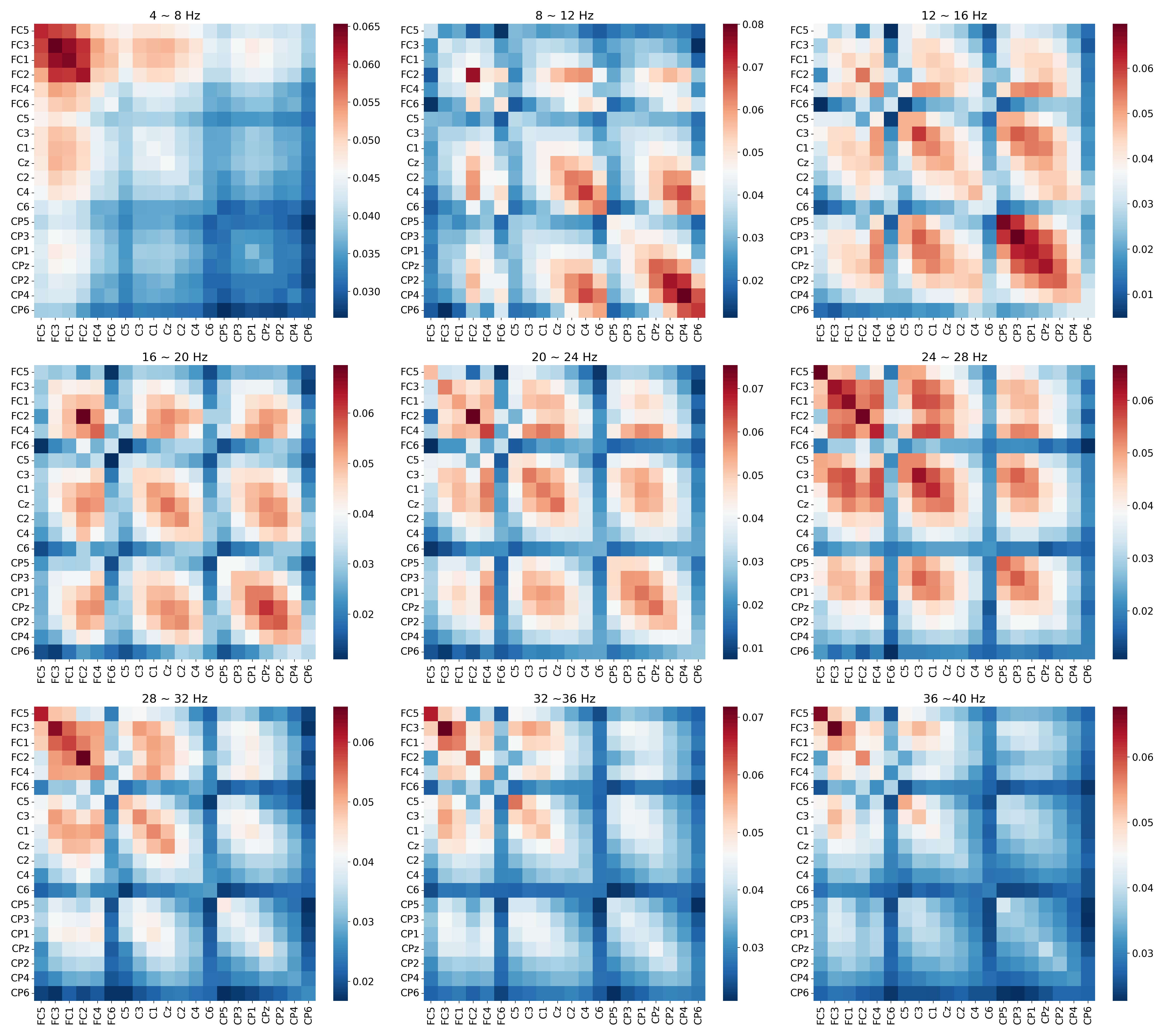}
\caption{Generated SCMs, \textbf{right}-hand (Sample 2)}~\label{fig:n}
\end{subfigure}
\caption{Conditional SCM Generation: 
In each line, we plot a picked SCM derived from an actual EEG segment for a category and another two generated samples within the same class. 
~\label{fig:3}
}
\end{figure*}

\section{EXPERIMENTS}

\subsection{The Korea University Dataset}
The Korea University (KU) Dataset~\footnote{\, The KU dataset refers to the \url{http://gigadb.org/dataset/100542} and its corresponding description article \url{https://academic.oup.com/gigascience/article/8/5/giz002/5304369}.}, also known as the OpenBMI dataset, was collected from 54 subjects performing a binary class EEG-MI task (i.e., the \textbf{left-hand} movement, and the \textbf{right-hand} movement). 
The EEG signals were captured at a rate of 1,000 Hz using 62 Ag/AgCl electrodes. 
The continuous EEG data were then segmented from 1,000 to 3,500 ms with reference to the stimulus onset. 
For evaluation, 20 electrodes in the motor cortex region were selected (i.e., FC-5/3/1/2/4/6, C-5/3/1/z/2/4/5, and CP-5/3/1/z/2/4/6). 
The study comprised two sessions, designated $S_1$ and $S_2$, each of which was divided into two phases, a training phase, and a test phase, each with 100 trials balanced between right and left-hand imagery tasks, resulting in a total of 21,600 trials (i.e., 54 subjects $\times$ 2 sessions $\times$ 200 trials) available for evaluation.

In accordance with the subject-specific study of the KU dataset in~\cite{ju2022graph}, the subjects were divided into two groups: a \emph{training subject group} and a \emph{test subject group}. 
The criteria for inclusion in the \emph{training subject group} were that the accuracies of 10-fold cross-validation on both $S_1$ and $S_2$ must be higher than 70\% (criterion level). 
A total of 21 subjects met this criterion and were included in the training subject group: Subjects No. 2, 3, 5, 6, 8, 12, 17, 18, 21, 22, 28, 29, 32, 33, 35, 36, 37, 39, 43, 44, and 45.
The remaining 33 subjects were included in the \emph{test subject group}: Subjects No. 1, 4, 7, 9, 10, 11, 13, 14, 15, 16, 19, 20, 23, 24, 25, 26, 27, 30, 31, 34, 38, 40, 41, 42, 46, 47, 48, 49, 50, 51, 52, 53, and 54.

\subsection{Parameters in Training Models}
For the score-based generative modeling, the Variance Exploding (VE) SDE approach, incorporating the NCSN++ model architecture~\footnote{\, The PyTorch implementation for Score-Based Generative Modeling refers to the GitHub repository (\url{https://github.com/yang-song/score_sde_pytorch}).}, was selected for evaluation.
We independently train two generative models to produce the left and right-hand samples using 4,200 trials in the \emph{training subject group}, comprising 21 subjects over 2 sessions with 100 trials/per class each. 
The signal from each trial was initially transformed into a covariance matrix and scaled. 
The noise parameters were set to $\sigma_{max} = 10$ and $\sigma_{min} = 0.01$. 
The training process was performed over 100,000 iterations, utilizing a batch size of 128. 
It is noteworthy that the CNN filter size within NCSN++ was set to $20 \times 20$ due to the reason discussed in Section~\ref{NonEuclidean}.
We pick $\epsilon = 1e-4$ in the projection step.

\subsection{Results}
To alleviate the discrepancy between the raw and generated distributions resulting from the generative methods, we normalize each covariance matrix by zero-centering the means and scaling the variances to unity before visualization and quantitative analysis.

\subsubsection{Visualization}
In Figure~\ref{fig:a}, we present 2-dimensional projections of the 8,400 trials' raw SCMs from the \emph{training subject group} and the generated 8,400 covariance matrices (both left and right-hand classes) using t-SNE~\cite{van2008visualizing}. 
The Fréchet means~\footnote{\, Fréchet mean is introduced in Appendix~\ref{A3}.} of both distributions are marked with triangle and cross signs, respectively. 
The two distributions are nearly coincident, and the Riemannian distance between their Fréchet means is relatively small (approximately 4.323), indicating that the center of the learned distribution closely matches the raw distribution. 
Figure~\ref{fig:b} provides a more detailed view of the projections within the Mu and Beta frequency bands. 
Furthermore, a set of covariance matrices corresponding to the two labeled Fréchet means in Figure~\ref{fig:a} have been plotted.
The Fréchet mean was computed independently for each frequency band. 
The Fréchet mean of the frequency bands $8\sim 12$ Hz, $12 \sim 16$ Hz, $16 \sim 20$ Hz, and $20 \sim 24$ Hz exhibit a similar profile, while the others differ. 
It is worth noting that these frequency bands are associated with event-related desynchronization and synchronization during cognitive and motor processing.
Figure~\ref{fig:2} illustrates the Fréchet means of covariance matrices differentiated between the left and right-hand classes within nine frequency bands.
Subfigures~\ref{fig:e} and~\ref{fig:f} display the SCMs with key regions highlighted, corresponding to neurophysiological findings.
Specifically, the highlighted entities in subfigures~\ref{fig:e} and~\ref{fig:g} (Mu and Beta bands) are situated in the regions of FC4, C4, and CP4 across the scalp, while those in subfigures~\ref{fig:f} and~\ref{fig:h} are located in the regions of FC3, C3, and CP3 across the scalp.
To provide a more comprehensive visual representation of the texture of the generated samples, Figure~\ref{fig:3} offers a closer examination of the SCMs derived from actual EEGs and those generated from the proposed methodology for the two categories.

\subsubsection{Classification}
To assess the generated samples' performance, we classify them using the pre-trained Tensor-CSPNet model~\footnote{\, The Python implementation of Tensor-CSPNet refers to the following GitHub repository: (\url{https://github.com/GeometricBCI/Tensor-CSPNet-and-Graph-CSPNet}).} on all subjects (two sessions) in the training subject group, which comprises a total of 8400 trials.
The model architecture adopts a simplified 2500 ms time window and incorporates two-level BiMap layers, transforming the input dimension of 20 to 30 and back to an output dimension of 20. 
There are 8400 balanced generated samples, with each class containing 4200. The pre-trained classifier predicts an accuracy of 84.30\% over all samples, and the confusion matrix is as follows:

\begin{table}[!ht]
\caption{Confusion Matrix: Predicted labels in a total of 8400. 
}
\centering 
\begin{tabular}{l | ll}
\toprule
True $\backslash$ Predicted & Right-hand & Left-hand \\
\midrule
Right-hand & 3730 (44.4\%) & 470 (5.6\%) \\
Left-hand & 849 (10.1\%) & 3351 (39.9\%)  \\
\bottomrule
\end{tabular}
\end{table}

\begin{table*}[!ht]
\caption{Cross-session classification with data augmentation approach: Each column depicts the number of samples incorporated into the training session. 
The samples are divided equally between the two classes - left-hand and right-hand. 
The selected cross-session scenario originates from the training and evaluation sessions in the KU dataset. 
The initial session of 200 trials and the added samples serve as the training data, while the first half of the second session, comprising 100 trials, are utilized for validation and the latter half, also consisting of 100 trials, for testing purposes. 
The results (\%) presented encompass the mean of \textbf{10} times runs across all scenarios and the optimal performance.
~\label{tab:cross_session}
}
\centering 
\begin{tabular}{l l | llll llllll}
\toprule
{Argumentation} & &&&& &&&&&\\
 {Samples} &  None & 20  & 40 & 60 & 80 & 100 & 120  & 140 &  160 & 180 & 200\\
\midrule
Subject No.30 & &&&& &&&& \\
Avg.(Std.) & 55.2(3.9) & 52.1(3.9) & 55.5(6.8)& 56.2(2.9)& 54.7(5.1) & 53.8(4.4) & 56.7(4.8)& 57.2(4.3)& 56.8(6.4)& \textbf{58.3}(5.3)& 57.6(4.5)\\
Best      & 61.0 & 59.0 & \textbf{71.0} & 63.0 & 64.0  & 62.0 & 64.0 & 66.0& 66.0& 66.0 & 67.0 \\
\midrule
Subject No.42 & &&&& &&&& \\
Avg.(Std.)  & 59.2(3.5) & 59.1(6.2) & 63.2(4.6)& 62.5(3.4)& 65.6(4.6)  & 65.1(4.1) & 64.8(3.6) & 65.8(2.8)& 65.4(3.6)& 61.8(4.1) & \textbf{67.9}(2.5) \\
Best     & 63.0 & 66.0 & 69.0& 67.0& 72.0 & \textbf{73.0} & 71.0 & 70.0& 72.0& 69.0 & 72.0\\
\bottomrule
\end{tabular}
\end{table*}

In this study, we conducted an additional experiment in a cross-session setting where one session of trials was utilized for training, the first half of another for validation, and the second half for testing, which is also known as the holdout scenario. 
This task presents a significant challenge due to the signal variability across sessions, and many state-of-the-art algorithms, including geometric methods~\cite{ju2022tensor, ju2022graph} performed poorly, yielding accuracy rates below 70\%. 
The proposed generative method was applied to generate SCMs using all subjects (two sessions) in the training subject group. 
The classifier, Tensor-CSPNet, was trained using the first session and the generated samples, validated in the first half of the second session, and evaluated in the second half for testing. 
Table~\ref{tab:cross_session} shows the cross-session classification accuracies, where each column of "None", "20", "40", "60", "80", "100", "120", "140", "160", "180", and "200" represents the number of added generative samples to the training set. The "None" column results are the typical cross-session outcomes but applied to normalized SCMs and without segmenting the time interval, thus slightly different from those in~\cite {ju2022graph}.
We selected Subjects 30 and 42 as representatives from the \emph{testing subject group}. In the case of Subject 30, the average accuracy, calculated over ten runs, increased by 3.1\% after the addition of 180 generated samples. 
Conversely, Subject 42 saw a substantial improvement of 8.7\% in average accuracy after incorporating 200 generated samples in each trial.

\subsection{Discussions}

This study explores a new method for generating SCMs for BCI applications using score-based generative modeling with the SDE approach. 
The generated samples are analyzed through both visual and quantitative evaluations. 
Visually, the samples produced by the proposed method have a comparable appearance to the SCMs obtained from actual EEG recordings. 
Furthermore, the center (Fréchet mean) of the generated samples aligns with neurophysiological findings that event-related desynchronization and synchronization occur on electrodes C3 and C4 within the Mu and Beta frequency bands during motor imagery processing. 
From a quantitative standpoint, 84.3\% of the samples can be accurately predicted by a pre-trained Tensor-CSPNet, and holdout experiments on two subjects (Subject No.30 and No.42) show an improvement of up to 8.7\% in the average accuracy of 10-times runs. 

Although our findings are promising, the absence of evaluation metrics for generative models in the EEG-BCI classification, akin to those commonly employed in the computer vision domain, such as Inception Score~\cite{salimans2016improved} and Fréchet Inception Distance~\cite{heusel2017gans}, precludes our study from providing more detailed outcomes, such as individual participant results. It is crucial to recognize that not all participants exhibit noticeable improvements after incorporating generative samples. At present, no established criterion exists for determining which subjects may benefit from this technique. Furthermore, the current approach has room for enhancement in several aspects, which we will address in following:

\subsubsection{Non-Euclidean Nature}~\label{NonEuclidean}
In the experiments, the SCM channels are ordered from start to end as FC-5/3/1/2/4/6, C-5/3/1/z/2/4/5, and CP-5/3/1/z/2/4/6. 
The score-based generative model employs a CNN-structured architecture to capture local information from adjacent channels in this sequence. 
However, this order fails to reflect the correlations between EEG channels with respect to their spatial locations, a phenomenon referred to as the non-Euclidean nature, which results in limited performance.
To tackle this problem, we propose a heuristic approach that sets the filter size to $20 \times 20$, which corresponds to the total size of the SCMs. 
It may not be readily applicable to complex scenarios, as it can be challenging to capture the signal granularity with a large filter size.

\subsubsection{Randomness}
It is possible that some of the generated samples may contain valuable discriminatory information for classification, while others may not. 
The randomness introduced by the sampling process in the score-based generative modeling may compromise the performance of the classifier.
Additionally, as we just mentioned, this randomness also leads to the ineffectiveness of conventional evaluation methods, which is due to the varying generated samples used for assessment each time, yet there is no common evaluation metric for the generative models in the EEG-BCI classification. 
After all, the texture of EEG spatial covariance matrices is not even present in the general image recognition databases.

\subsubsection{Cross-frequency Coupling}
A potential explanation for the limited performance could be attributed to the diversity of the generative model. 
Since each SCM over a specific frequency band is independently generated from random noise, the composite SCM generated from these independent SCMs may lack neurophysiological significance and have yet to be previously observed.
In simpler terms, real SCMs derived from the EEG signal where changes in brain activity occur during cognitive and motor processing, resulting in event-related desynchronization and synchronization. 
However, a generative SCM may not have this same origin, even though it may appear similar.
For instance, the SCM within the frequency range of 32 to 36 Hz, as depicted in Subfigure~\ref{fig:m}, highlights a novel instance of the typical occurrence of high-intensity activities within the Mu and Beta bands.

\subsubsection{Distribution Shift}
Despite the fact that the generative samples may contain ample discriminatory information, the limited performance observed may still stem from the disparity between the prior and learned distributions. 
This incongruence can result in variations in the numerical ranges of pixels or entities within the SCMs. 
To mitigate this challenge, we utilize a simple heuristic normalization technique for the covariance matrices by zero-centering the means and scaling the variances to unity. 
This approach results in well-overlapping raw and generated distributions, but it may not always be a reliable method in complex scenarios.

\section{APPENDICES}

In the appendices, we will briefly introduce score-based generative modeling. 
For a formal convention, we suppose we have samples of spatial covariance matrices $\{S_i \in \mathbb{R}^{n_C \times n_C}\}_{i=1}^N$ from an (unknown) distribution $p_{data}(S)$.
\subsection{Score-based Generative Modeling}~\label{A1}
In the score-based generative modeling, the score network $s_{\theta}: \mathbb{R}^{n_C \times n_C} \longmapsto \mathbb{R}^{n_C \times n_C}$ is a deep neural network parametrized by $\theta$ and used to learn the score of a probability density $\nabla_S \log p(S)$. 
To train score network $s_{\theta}$, a technique called \emph{denoising score matching}~\cite{vincent2011connection} is proposed to firstly replace $p_{data}(S)$ using a Gaussian noise $\sigma$-pertubed version $p_{data}^{\sigma}(\tilde{S})$, where $p_{data}^{\sigma}(\tilde{S}) = \int p_{\mathcal{N}}^{\sigma}(\tilde{S}|S) \cdot p_{data} (S)~dS$, and the denoising objective $\mathcal{J}_D (\theta)$ with noise level $\sigma$ is then given as follows,
\[
\mathcal{J}_D^{\sigma} (\theta) := \mathbb{E}_{p_{\mathcal{N}}^{\sigma}(\tilde{S}|S) \cdot p_{data} (S)} ||s_{\theta}(\tilde{S}) - \nabla_{\tilde{S}} \log p_{\mathcal{N}}^{\sigma}(\tilde{S}|S)||. 
\]
Keep in mind that the noise model term $\nabla_{\tilde{S}} \log p_{\mathcal{N}}^{\sigma}(\tilde{S}|S)$ has a simple analytic form, written $\nabla_{\tilde{S}} \log p_{\mathcal{N}}^{\sigma}(\tilde{S}|S) = -\frac{1}{\sigma^2} (\tilde{S} - S)$.
In the sampling phase, Langevin dynamics are applied to recursively generate samples using the score function $s_{\theta}$ as follows, 
\[
\tilde{S}_t = \tilde{S}_{t-1} + \frac{\epsilon}{2}\cdot s_{\theta} (\tilde{S}_{t-1}) + \sqrt{\epsilon}\cdot Z_t,
\]
where initial $\tilde{S}_0 \sim \pi(x)$ (prior distribution) and fixed step size $\epsilon > 0$ are given, and $Z_t \in \mathcal{N} (0, I)$.
\subsection{Diffusing Samples with an SDE}~\label{A2}
For a continuum of distribution evolving over time $t$, the score-based generative modeling has been further established within a unified framework of stochastic diffusion equations (SDEs) with diffusion probabilistic modeling.~\cite{song2021scorebased}
Technically, the SDEs are of the form as follows, 
\[
dS = f(S, t)~dt + g(t)~dW, 
\]
where $f(\cdot, t): \mathbb{R}^{n_C \times n_C} \longmapsto \mathbb{R}^{n_C \times n_C}$ and $g(t) \in \mathbb{R}$ are the drift and diffusion coefficient respectively, and $W\in \mathbb{R}^{n_C \times n_C}$ is a standard Wiener process.
The solution of the above SDE is a diffusion process $\{S(t)\}_{t \in [0, T]}$ over a finite time horizon $[0, T]$, and $p_t(S)$ is the marginal distribution of $S(t)$.
For variance exploding (VE) SDEs, $f(S_t, t) := \alpha_t \cdot s_\theta (S_t)$ and $g(S_t, t) := \sqrt{2\alpha_t}$.
The score-based generative modeling relies on the following time-reversal diffusion process for generating samples, 
\[
d S = \big( f(x, t) - g(t)^2 \cdot \nabla_S \log p_t(S) \big)~dt + g(t)~d \bar{W}, 
\]
where $\bar{W}\in \mathbb{R}^{n_C \times n_C}$ is a standard Wiener process in the reverse-time direction. 
A time-dependent neural network $s_{\theta} (S, t)$ is used to estimate $\nabla_S \log p_t(S)$ by squeezing the following loss $\mathcal{J}_D \big(\theta; \lambda\big)$ as
\[
\int_0^T \mathbb{E}_{p_{0t}(\tilde{S}|S) \cdot p_{data} (S)} \lambda(t) \cdot ||s_{\theta}(\tilde{S}, t) - \nabla_{\tilde{S}} \log p_{0t} (\tilde{S}|S)||~dt, 
\]
where $p_{0t} (\tilde{S}|S)$ is the transition distribution from $S(0)$ to $S(t)$, and $\lambda:[0, T] \rightarrow \mathbb{R}_{>0}$ is a positive weighting function.

\subsection{Fréchet Mean}~\label{A3}
The generated sample in the current approach is in the form of an SPD matrix, which means that traditional measures for generative modeling in computer vision, such as the Inception score and Fréchet Inception Distance, cannot be used. 
The Riemannian distance on SPD manifolds is used as an alternative method to evaluate the distance between the prior and generated distributions.
From a mathematical perspective, a conventional treatment to view the space of spatial covariance matrices is on the symmetric positive definite (SPD) manifolds, which is equipped with a Riemannian metric, i.e., affine invariant Riemannian metric (AIRM)~\cite{pennec2006riemannian}, written as $(\mathcal{S}^{++}, AIRM)$.
The Riemannian distance between two spatial covariance matrices is $d_{AIRM} (S_1, S_2) := ||\log{(S_1^{-1} \cdot S_2)}||_{\mathcal{F}}$, where $\mathcal{F}$ is Frobenius norm and $\log$ is the logarithm. 
Given a set of SPD matrices $\{S^1, \cdots, S^N\}$, the Fréchet mean $\mu$ of that set is given as follows, 
\[
\mu := \arg\min_{\mu \in \mathcal{S}^{++}} \frac{1}{N} \cdot \sum_{i=1}^N~d_{AIRM}^2 (S^i, \mu).
\]

\subsection{Mathematical Fundamentals in the Projection Step}~\label{sec:projection}
Consider a $d \times d$ real symmetric matrix $S$ that possesses eigenvalues $\lambda_1 \geq \lambda_2 \geq \cdots \geq \lambda_d$ and corresponding orthonormal eigenvectors $\{u_i\}_{i=1}^d$. In this context, the spectral decomposition of $S$ is expressed as $S = \sum_{i=1}^d \lambda_i \cdot u_i u_i^T$.
To deal with nonnegative eigenvalues in symmetric matrices, the following lemma from~\cite{ju2022graph} introduces a fundamental technique: 
Projection $S^\dag:=\sum_{i=1}^d \max{\{\lambda_i, 0\}} \, u_i u_i^T$ on the positive semidefinite cone is the extremum of the minimization problem $||S-S^\dag||_2^2$ subject to $S \succeq 0$.

\section{ACKNOWLEDGMENT}
This work was supported under the RIE2020 Industry Alignment Fund–Industry Collaboration Projects (IAF-ICP) Funding Initiative, as well as cash and in-kind contributions
from industry partner(s);
This work was supported by the RIE2020 AME Programmatic Fund, Singapore (No. A20G8b0102);
This work was also supported by Innovative Science and Technology Initiative for Security Grant Number JPJ004596, ATLA, Japan.

{\footnotesize
\bibliographystyle{IEEEtran}
\bibliography{refs}

\begin{thebibliography}{10}
\providecommand{\url}[1]{#1}
\csname url@samestyle\endcsname
\providecommand{\newblock}{\relax}
\providecommand{\bibinfo}[2]{#2}
\providecommand{\BIBentrySTDinterwordspacing}{\spaceskip=0pt\relax}
\providecommand{\BIBentryALTinterwordstretchfactor}{4}
\providecommand{\BIBentryALTinterwordspacing}{\spaceskip=\fontdimen2\font plus
\BIBentryALTinterwordstretchfactor\fontdimen3\font minus
  \fontdimen4\font\relax}
\providecommand{\BIBforeignlanguage}[2]{{%
\expandafter\ifx\csname l@#1\endcsname\relax
\typeout{** WARNING: IEEEtran.bst: No hyphenation pattern has been}%
\typeout{** loaded for the language `#1'. Using the pattern for}%
\typeout{** the default language instead.}%
\else
\language=\csname l@#1\endcsname
\fi
#2}}
\providecommand{\BIBdecl}{\relax}
\BIBdecl

\bibitem{schirrmeister2017deep}
R.~T. Schirrmeister, J.~T. Springenberg, L.~D.~J. Fiederer, M.~Glasstetter,
  K.~Eggensperger, M.~Tangermann, F.~Hutter, W.~Burgard, and T.~Ball, ``Deep
  learning with convolutional neural networks for eeg decoding and
  visualization,'' \emph{Human brain mapping}, vol.~38, no.~11, pp. 5391--5420,
  2017.

\bibitem{ju2020federated}
C.~Ju, D.~Gao, R.~Mane, B.~Tan, Y.~Liu, and C.~Guan, ``Federated transfer
  learning for eeg signal classification,'' in \emph{2020 42nd Annual
  International Conference of the IEEE Engineering in Medicine \& Biology
  Society (EMBC)}.\hskip 1em plus 0.5em minus 0.4em\relax IEEE, 2020, pp.
  3040--3045.

\bibitem{hartmann2018eeg}
K.~G. Hartmann, R.~T. Schirrmeister, and T.~Ball, ``Eeg-gan: Generative
  adversarial networks for electroencephalograhic (eeg) brain signals,''
  \emph{arXiv preprint arXiv:1806.01875}, 2018.

\bibitem{ju2022tensor}
C.~Ju and C.~Guan, ``Tensor-cspnet: A novel geometric deep learning framework
  for motor imagery classification,'' \emph{IEEE Transactions on Neural
  Networks and Learning Systems}, 2022.

\bibitem{ju2022deep}
------, ``Deep optimal transport on spd manifolds for domain adaptation,''
  \emph{arXiv preprint arXiv:2201.05745}, 2022.

\bibitem{ju2022graph}
------, ``Graph neural networks on spd manifolds for motor imagery
  classification: A perspective from the time-frequency analysis,'' \emph{arXiv
  preprint arXiv:2211.02641}, 2022.

\bibitem{kobler_neurips22}
R.~Kobler, J.-i. Hirayama, Q.~Zhao, and M.~Kawanabe, ``Spd domain-specific
  batch normalization to crack interpretable unsupervised domain adaptation in
  eeg,'' in \emph{Advances in Neural Information Processing Systems}, vol.~35,
  2022, pp. 6219--6235.

\bibitem{pan2022matt}
Y.-T. Pan, J.-L. Chou, and C.-S. Wei, ``Matt: A manifold attention network for
  eeg decoding,'' \emph{arXiv preprint arXiv:2210.01986}, 2022.

\bibitem{wilson2022deep}
D.~Wilson, L.~A.~W. Gemein, R.~T. Schirrmeister, and T.~Ball, ``Deep riemannian
  networks for eeg decoding,'' \emph{arXiv preprint arXiv:2212.10426}, 2022.

\bibitem{song2019generative}
Y.~Song and S.~Ermon, ``Generative modeling by estimating gradients of the data
  distribution,'' \emph{Advances in Neural Information Processing Systems},
  vol.~32, 2019.

\bibitem{song2020improved}
------, ``Improved techniques for training score-based generative models,''
  \emph{Advances in neural information processing systems}, vol.~33, pp.
  12\,438--12\,448, 2020.

\bibitem{song2021scorebased}
Y.~Song, J.~Sohl-Dickstein, D.~P. Kingma, A.~Kumar, S.~Ermon, and B.~Poole,
  ``Score-based generative modeling through stochastic differential
  equations,'' in \emph{International Conference on Learning Representations},
  2021.

\bibitem{van2008visualizing}
L.~Van~der Maaten and G.~Hinton, ``Visualizing data using t-sne.''
  \emph{Journal of machine learning research}, vol.~9, no.~11, 2008.

\bibitem{salimans2016improved}
T.~Salimans, I.~Goodfellow, W.~Zaremba, V.~Cheung, A.~Radford, and X.~Chen,
  ``Improved techniques for training gans,'' \emph{Advances in neural
  information processing systems}, vol.~29, 2016.

\bibitem{heusel2017gans}
M.~Heusel, H.~Ramsauer, T.~Unterthiner, B.~Nessler, and S.~Hochreiter, ``Gans
  trained by a two time-scale update rule converge to a local nash
  equilibrium,'' \emph{Advances in neural information processing systems},
  vol.~30, 2017.

\bibitem{vincent2011connection}
P.~Vincent, ``A connection between score matching and denoising autoencoders,''
  \emph{Neural computation}, vol.~23, no.~7, pp. 1661--1674, 2011.

\bibitem{pennec2006riemannian}
X.~Pennec, P.~Fillard, and N.~Ayache, ``A riemannian framework for tensor
  computing,'' \emph{International Journal of computer vision}, vol.~66, no.~1,
  pp. 41--66, 2006.

\end{thebibliography}
}

\end{document}